\documentclass[aps,pra,twocolumn,superscriptaddress]{revtex4}
\usepackage{amsfonts,amssymb,graphicx}
\usepackage{amsmath}
\usepackage[usenames,dvipsnames]{xcolor}
\usepackage{pdfpages}
\usepackage{footmisc}
\usepackage[normalem]{ulem}
\usepackage{color}

\begin{document}

\title{Relaxation of antiferromagnetic order and growth of R\'enyi entropy in a generalized Heisenberg star}
\author{Jiaxiu Li}
\affiliation{Center for Quantum Technology Research, School of Physics, Beijing Institute of Technology, Beijing 100081, China}
\affiliation{Key Laboratory of Advanced Optoelectronic Quantum Architecture and Measurements (MOE), School of Physics, Beijing Institute of Technology, Beijing 100081, China}
\author{Ye Cao}
\affiliation{Key Laboratory of Advanced Optoelectronic Quantum Architecture and Measurements (MOE), School of Physics, Beijing Institute of Technology, Beijing 100081, China}
\author{Ning Wu}
\email{wunwyz@gmail.com}
\affiliation{Center for Quantum Technology Research, School of Physics, Beijing Institute of Technology, Beijing 100081, China}
\affiliation{Key Laboratory of Advanced Optoelectronic Quantum Architecture and Measurements (MOE), School of Physics, Beijing Institute of Technology, Beijing 100081, China}
\begin{abstract}
Interacting central spin systems, in which a central spin is coupled to a strongly correlated spin bath with intrabath interaction, consist of an important class of spin systems beyond the usual Gaudin magnet. These systems are relevant to several realistic setups and serve as an interesting platform to study interaction controlled decoherence and frustration induced instability of magnetic order. Using an equations-of-motion method based on analytical representations of spin-operator matrix elements in the XX chain, we obtain exact long-time dynamics of a generalized Heisenberg star consisting of a spin-$S$ central spin and an inhomogeneously coupled XXZ chain of $N\leq 16$ bath spins. In contrast to previous studies where the central spin dynamics is mainly concerned, we focus on the influence of the central spin on the dynamics of magnetic orders within the spin bath. By preparing the XXZ bath in a N\'eel state, we find that in the gapless phase of the bath even weak system-bath coupling could lead to nearly perfect relaxation of the antiferromagnetic order. In the gapped phase, the staggered magnetization decays rapidly and approaches a steady value that increases with increasing anisotropy parameter. These findings suggest the possibility of controlling internal dynamics of strongly correlated many-spin systems by certain coupled auxiliary systems of even few degrees of freedom. We also study the dynamics of the R\'enyi entanglement entropy of the central spin when the bath is prepared in the ground state. Both the overall profile and initial growth rate of the R\'enyi entropy are found to exhibit minima at the critical point of the XXZ bath.
\end{abstract}

\maketitle

\section{Introduction}
\par Quantum spin systems serve as a paradigm exhibiting strong correlations and many-body effects and their ground-state properties have long attracted considerable interest since the early work by Bethe in the 1930s~\cite{Bethe}. Theoretical investigations of many-body spin systems are usually challenging due to complex interactions and exponential growth of the Hilbert space dimension with the system size. In this context exactly soluble models provide a useful guide to understand general properties of the ground state, excited states, and nonequilibrium dynamics of more general quantum spin systems.
\par Two typical classes of widely studied soluble models are quantum spin chains (e.g., the quantum XY chain, the XXZ chain) and Gaudin-like models (e.g., the Gaudin magnet or central spin model, the reduced BCS model), both of which can be solved by certain types of Bethe ansatz or free-fermion techniques~\cite{Takbook,Gaudin1976,RMP2004}. Nearly thirty years ago, Richter and Voigt proposed a simple frustrated spin model combining the above two ingredients and named it as a Heisenberg star (referred to as the R-V model below)~\cite{Heisenbergstar}. The R-V model consists of a spin-1/2 central spin and a homogeneously coupled spin bath modeled by an XXX chain~\cite{Heisenbergstar},
\begin{eqnarray}\label{RV1}
H_{\mathrm{R-V}}=J\sum^N_{j=1}\vec{S}_j\cdot\vec{S}_{j+1}+2g\sum^N_{j=1}\vec{S}\cdot\vec{S}_j,~J,~g>0
\end{eqnarray}
where $J$ and $2g$ are, respectively, the intrabath and system-bath coupling strengths. The competition between the two terms is found to result in interesting behaviors of ground-state spin correlations~\cite{Heisenbergstar}.
\par In its original form, the central spin model consists of a central spin and an inhomogeneously coupled spin bath without intrabath coupling and is integrable under certain conditions~\cite{Gaudin1976,Kiss2001,NPB2005,Claeys2015,Physica2018,PRresearch2020,PRB2020}. The central spin model and its variants have attracted great attention in the past decades due to their relevance to quantum decoherence~\cite{Loss2002,Raedt2004,Raedt2005,PRA2005,Raedt2008,Stolze2010,Sarma2011,Sarma2012,Coish2020,Uhrig2013,PRL2013,DSPRB2013}, quantum information~\cite{Dooley2013,PRA2020}, quantum metrology~\cite{Guan2019}, and quantum batteries~\cite{GuanQB,YangQB}. Since the spin bath is itself noninteracting and featureless, theoretical studies of such kind of central spin systems mainly focus on the bath induced central spin dynamics, which has been extensively studied using various methods, including quantum master equations~\cite{Sarma2012}, Bethe ansatz based techniques~\cite{PRL2013,DSPRB2013,Guan2019}, and the time-dependent density matrix renormalization group (t-DMRG)~\cite{Uhrig2013}, etc.
\par However, in more realistic cases the intrabath coupling between bath spins cannot be neglected and will affect the long-time dynamics of the system. These central spin systems with self-interacting spin baths will be referred to as interacting central spin models (ICSMs) and their dynamics has been widely investigated~\cite{JPA2002,Quan2006,Sun2007,Cao2007,Fan2007,PRA2007,Yuan2007,Paz2007,Zou2010,Liu2009,OC2009,Hu2010,Chen2008,NC2014,PRA2014,PRB2016,Yang2020,Tanimura2021,CTP2022}. In the context of spin baths modeled by one-dimensional spin chains, the central spin dynamics affected by a variety of structured spin environments, including the quantum Ising chain~\cite{Quan2006,Sun2007,Cao2007}, the XY chain~\cite{Fan2007,PRA2007,Yuan2007,Paz2007,Zou2010}, and spin chains with Dzyaloshinskii-Moriya interactions~\cite{Liu2009,OC2009,Hu2010}, have been extensively studied. There are several common features shared by these studies: i) The spin baths considered in these works can be mapped onto free-fermion models, ii) Only non-spin-flipping or dephasing interactions between the central spin and the spin chain are considered, iii) The central spin dynamics (e.g., decoherence or entanglement dynamics) influenced by the criticality of the spin bath is mainly concerned.
\par Contrastingly, it is more difficult to evaluate the long time dynamics of ICSMs in which the spin-flipping system-bath coupling or strongly correlated terms in the spin bath (e.g., the $zz$ interaction) are introduced. In the former case, the decoherence dynamics of a qubit coupled to an XX chain via XX-type~\cite{PRA2014} and XXZ-type~\cite{PRB2016} system-bath coupling is studied by using an equations-of-motion method combined with the Chebyshev expansion technique. The spin-flipping interaction makes the free-fermion solution of the spin bath inapplicable due to the nonlocal nature of the spin raising and lowering operators in the fermion representation. There appeared only a few works in which strongly correlated spin baths were addressed. The decoherence and entanglement dynamics of a single qubit or two qubits coupled to an XXZ chain through Ising-type~\cite{PRA2007} or local XXX-type~\cite{Chen2008} interactions are studied using the t-DMRG. The central spin coherence and polarization dynamics in a homogeneous Heisenberg star with a Zeeman term is studied based on the Bethe ansatz solutions of the XXX bath~\cite{Yang2020}. The finite-temperature decoherence dynamics of a two-level system interacting with an XXZ chain is studied using the hierarchical equations of motion method~\cite{Tanimura2021}. Most recently, the R-V model is generalized to the case of a higher central spin and its ground-state and dynamical properties are rigorously investigated through analytical and numerical approaches~\cite{CTP2022}.
\par In spite of the above-mentioned works, the influence of the system part, the central spin, on the dynamical behaviors of the strongly correlated spin bath is largely unexplored. As perhaps the first ICSM that combines the Gaudin magnet and a nontrivial spin bath, the R-V model strongly suggest us to investigate the effects of the central-spin induced frustration on the internal properties of the bath. For example, it is demonstrated in Ref.~\cite{NC2014} that the central spin decoherence can be used as a tool to detect many-body correlations in the coupled spin environment. The study of central-spin driven dynamics of the spin bath will help us gain a better understanding of the influence of small quantum systems of few degrees of freedom on the nonequilibrium dynamics of strongly correlated many-body systems.
\par In this work, we study the real-time dynamics of an ICSM that is a generalization of the R-V model (referred to as a generalized Heisenberg star), where the spin bath is modelled by a spin-1/2 antiferromagnetic XXZ chain inhomogeneously coupled to a spin-$S$ central spin via XXZ-type system-bath coupling. The XXZ chain is a paradigmatic strongly correlated spin model and its dynamical properties continue to attract the attention of theorists~\cite{PRL2009,NJP2010,Caux2010,Heyl2014,Sirker2021}. This is mainly motivated by experimental advances in cold-atom systems, where the spin-1/2 and spin-1 XXZ chains have been realized and certain initial states are successfully prepared~\cite{Nature2013,Nature2020,Ketterle2021}. To be specific, we prepare the XXZ bath in a N\'eel state and investigate the relaxation of the antiferromagnetic order driven simultaneously by the intrabath nearest-neighbor antiferromagnetic interaction and the system-bath coupling. We employ an equations-of-motion method~\cite{PRA2014,PRB2016} to treat the dynamics of the whole system with $N\leq 16$ bath spins. The usefulness of the method lies in the fact that each bath spin interacts locally to the central spin, while the matrix elements of bath operators in the diagonal basis of the XX chain admit analytical expressions~\cite{SOME2018}. By numerically solving the equations of motion in each magnetization sector, we are able to obtain the exact dynamics of the generalized Heisenberg star prepared in a generic uncorrelated initial state.
\par We find that even weak system-bath coupling can yield nearly perfect relaxation of the antiferromagnetic order in the gapless phase of the XXZ bath. In the gapped phase of the XXZ bath, the staggered magnetization rapidly approaches a finite steady value for strong system-bath couplings. However, at the critical point of the XXZ bath the staggered magnetization keeps an oscillatory behavior around zero value from the weak to strong system-bath couplings. These observations are in sharp contrast to the case of vanishing system-bath coupling~\cite{PRL2009,NJP2010} and can be qualitatively understood from an energetic point of view and by looking at the corresponding decoherence of the central spin. We also find that for strong system-bath couplings increasing the quantum number $S$ of the central spin further facilitates the initial decay and suppresses the steady value of the staggered magnetization. These findings indicate that even a simple quantum system with few degrees of freedom could have significant influence on the dynamics of the coupled many-body system.
\par Our method also allows us to study the reduced dynamics of the central spin in the usual way. In this case we prepare the XXZ bath in its ground state and focus on the growth of the R\'enyi entanglement entropy of the central spin as a measure of entanglement between the central spin and the bath. Remarkably, we find that the R\'enyi entanglement entropy acquires the lowest value at the critical point of the XXZ bath. The short-time dynamics is found to be of a Gaussian form, which basically reflects the overall behavior of the entropy on longer time scales. Correspondingly, the short-time growth rate also achieves a minimum at the critical point. The critical properties of the XXZ bath can thus be detected through the entanglement dynamics of the central spin, providing a way to investigate the internal phases of strongly correlated spin systems via probing the corresponding simple auxiliary system with which it interacts.
\par The rest of the paper is organized as follows. In Sec.~\ref{SecII} we introduce the generalized Heisenberg star and provide details of the equations-of-motion approach based on spin-operator matrix elements of the XX chain. In Sec.~\ref{SecIII} we study the relaxation of antiferromagnetic order measured by the staggered magnetization when the XXZ bath is prepared in the N\'eel state. In Sec.~\ref{SecIV} we study the dynamics of the R\'enyi entanglement entropy of a higher central spin when the XXZ bath is prepared in its ground state.  Conclusions are drawn in Sec.~\ref{SecV}.
\section{Model and Methodology}\label{SecII}
\subsection{Hamiltonian}
\begin{figure}
\includegraphics[width=.50\textwidth]{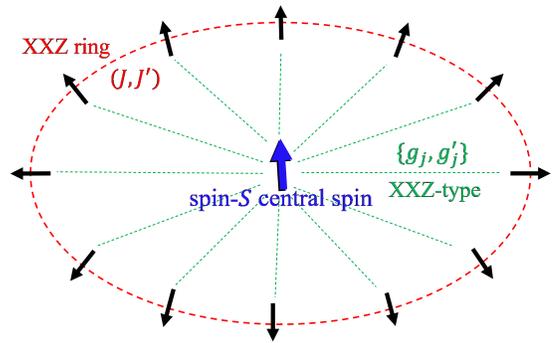}
\caption{An inhomogeneous Heisenberg star consists of a spin-$S$ central spin and a spin bath modeled by an XXZ ring, with the two part interacting via inhomogeneous XXZ-type hyperfine coupling.}
\label{Fig1}
\end{figure}

We consider a generalized inhomogeneous Heisenberg star described by the Hamiltonian (see Fig.~\ref{Fig1})
\begin{eqnarray}
H&=&H_{\mathrm{S}}+H_{\mathrm{B}}+H_{\mathrm{SB}}.
\end{eqnarray}
The system part
\begin{eqnarray}
H_{\mathrm{S}}&=&\omega  S_z+\lambda S^2_z,
\end{eqnarray}
describes a central spin $\vec{S}=(S_x,S_y,S_z)$ of size $S\geq1/2$, where $\omega$ is the Larmor frequency due to the applied magnetic field and $\lambda$ is the single-ion anisotropy of the central spin when $S\geq 1$. The spin bath takes the form of a spin-1/2 XXZ chain
\begin{eqnarray}
H_{\mathrm{B}}&=&H_{\mathrm{XX}}+H_{\mathrm{Z}},\nonumber\\
H_{\mathrm{XX}}&=&J\sum^N_{j=1}(S^x_jS^x_{j+1}+S^y_jS^y_{j+1}),\nonumber\\
H_{\mathrm{Z}}&=&J'\sum^N_{j=1} S^z_jS^z_{j+1},
\end{eqnarray}
where $\vec{S}_j=(S^x_j,S^y_j,S^z_j)$ is the spin-1/2 operator for the $j$th bath spin. We have separated the bath Hamiltonian into the in-plane component $H_{\mathrm{XX}}$ and the Ising component $H_{\mathrm{Z}}$. For simplicity, we assume that $N$ is even and impose periodic boundary conditions. We set $J>0$ and the sign of $J'$ essentially determines the quantum phase of $H_{\mathrm{B}}$~\cite{Nagaosa}. The XXZ-type hypefine coupling between the central spin and the spin bath reads
\begin{eqnarray}
H_{\mathrm{SB}}&=&2\sum^N_{j=1}[g_j(S_xS^x_j+S_yS^y_j)+g'_jS_zS^z_j]\nonumber\\
&=& \sum^N_{j=1}[g_j(S_+S^-_j+S_-S^+_j)+2g'_jS_zS^z_j],
\end{eqnarray}
where $\{g_j\}$ and $\{g'_j\}$ are, respectively, the in-plane and Ising parts of the (inhomogeneous) exchange interaction constants. It is usually the case that $g'_j/g_j=\Lambda,~\forall j$, where $\Lambda$ measures the anisotropy of the hyperfine coupling. The Heisenberg star $H$ might be simulated in cold-atom systems by engineering the interaction between an XXZ chain with an auxiliary central atom. It can also describe the physics of one-dimensional molecular aggregates strongly coupled to a microcavity, where the molecular aggregate is modeled by a Frenkel exciton model with exciton-exciton interaction~\cite{Ezaki1994,SpanoCPL1995} and the few-photon states are mimicked by the spin-$S$ central spin~\cite{LPP2016}.
\par Let $\vec{L}\equiv\sum^N_{j=1}\vec{S}_j$ be the collective angular momentum operator of the spin bath, it can be easily checked that the total magnetization $\hat{M}=S_z+L_z$ is conserved. The angular momentum of the central spin $\vec{S}^2$ is also conserved. However, the total angular momentum of the spin bath, $\vec{L}^2$, is not conserved unless $J=J'$ and $\{g_j\}$ and $\{g'_j\}$ are both homogeneous~\cite{PRB2016}. Below the eigenvalues of $\hat{M}$, $S_z$, and $L_z$ will be denoted as $M$, $s_z$, and $l_z$, respectively.
\par In the case of $J=J'=0$, the bath becomes noninteracting and we recover the usual Gaudin magnet that admits Bethe ansatz solutions under certain conditions~\cite{Gaudin1976,Kiss2001,NPB2005,Claeys2015,Physica2018,PRresearch2020,PRB2020}. In the case of $J'=0$ and $S=1/2$, the generalized Heisenberg star is reduced to the XX star studied in Ref.~\cite{PRB2016}. In the special case of $\omega=\lambda=0$, $J=J'$, and $g_j=g'_j=g,~\forall j$, the Hamiltonian is reduced to the R-V model given by Eq.~(\ref{RV1}), which can be rewritten as
\begin{eqnarray}\label{RV2}
H_{\mathrm{R-V}}=J\sum^N_{j=1}\vec{S}_j\cdot\vec{S}_{j+1}+ g(\vec{\mathcal{J}}^2-\vec{S}^2-\vec{L}^2),
\end{eqnarray}
where $\vec{\mathcal{J}}=\vec{L}+\vec{S}$ is the total angular momentum of the system. The R-V model possesses a number of symmetries, and hence conserves the following quantities, i.e., the total energy $H_{\mathrm{R-V}}$, the total angular momentum $\vec{\mathcal{J}}^2$, the total magnetization $\hat{M}$, the bath angular momentum $\vec{L}^2$, and the bath energy $H_{\mathrm{B}}=J\sum^N_{j=1}\vec{S}_j\cdot\vec{S}_{j+1}$. As a result, a general eigenstate of $H_{\mathrm{R-V}}$ can be labeled by, respectively, the corresponding quantum numbers as $|\psi_{E,\mathcal{J},M,l}\rangle$.
\par The total magnetization $M$ takes the following $2S+N+1$ possible values: $M=-S-\frac{N}{2},-S-\frac{N}{2}+1,\cdots,S+\frac{N}{2}$. The structure of the states in an individual $M$-subspace depends on whether $S<\frac{N}{2}$ or $S\geq\frac{N}{2}$. In this paper, we focus on the case of $S< \frac{N}{2}$ (see Appendix~\ref{AppA}). To get a universal short-time dynamics for different numbers of bath spins, we introduce the energy scale
\begin{eqnarray}
\omega_{\mathrm{fluc}}&=&2\sqrt{\sum^N_{j=1}g^2_j},
\end{eqnarray}
which is associated with the fluctuation of the Overhauser field~\cite{OverF}.
\subsection{Method: spin-operator matrix elements}
\par To numerically simulate the real-time dynamics of the composite system, we use the representation in which the Hamiltonian $H_0=H_{\mathrm{S}}+H_{\mathrm{XX}}$ is diagonal. This is motivated by the fact that the matrix elements of each term in the remaining part of the Hamiltonian, $H_1=H-H_0$, can be expressed in this representation in terms of the so-called spin-operator matrix elements for the XX chain~\cite{SOME2018}. The eigenbasis of $H_0$ is spanned by the following $(2S+1)2^N$ states
\begin{eqnarray}
\{|s_z\rangle|\vec{\eta}_n\rangle\},~s_z=S,S-1,\cdots,-S;~n=0,1,\cdots,N,\nonumber
\end{eqnarray}
where
\begin{eqnarray}
S_z|s_z\rangle&=&s_z|s_z\rangle,\nonumber\\
H_{\mathrm{XX}}|\vec{\eta}_n\rangle&=&\mathcal{E}_{\vec{\eta}_n}|\vec{\eta}_n\rangle,
\end{eqnarray}
with $\mathcal{E}_{\vec{\eta}_n}=\sum^n_{l=1} J\cos K^{(\sigma_n)}_{\eta_l}$. Here, $|\vec{\eta}_n\rangle$ is an eigenstate of $H_{\mathrm{XX}}$ having $n$ fermionic excitations labeled by the tuple $\vec{\eta}_n=(\eta_1,\cdots,\eta_n)$  (with the convention $1\leq\eta_1<\cdots<\eta_n\leq N$) with respect to the vacuum state $|0\rangle=|\downarrow\cdots\downarrow\rangle$~\cite{SOME2018}. The corresponding eigenenergy $\mathcal{E}_{\vec{\eta}_n}$ depends on the parity of $n$ through wave numbers $
K^{(\sigma_n)}_{\eta_l}=-\pi+\left[2\eta_l+\frac{1}{2}(\sigma_n-3)\right]\frac{\pi}{N}$, with $\sigma_n=1$ (even $n$) or $\sigma_n=-1$ (odd $n$). For later convenience, we also introduce $\alpha=s_z+n=M+\frac{N}{2}$, which is also conserved and takes values from $\alpha=-S$ to $\alpha=S+N$.
\par As we will see, the equations of motion of the system in the basis $\{|s_z\rangle|\vec{\eta}_n\rangle\}$ involve the following matrix elements
\begin{eqnarray}
F_{\vec{\eta}_{n+1},\vec{\chi}_n}(\{g_j\})&=&\langle\vec{\chi}_n|\sum^N_{j=1}g_jS^-_j|\vec{\eta}_{n+1}\rangle,\nonumber\\
G_{\vec{\chi}_{n},\vec{\chi}'_n}(\{g'_j\})&=&\langle\vec{\chi}_n|\sum^N_{j=1}g'_jS^z_j|\vec{\chi}'_{n}\rangle,\nonumber\\
\bar{G}_{\vec{\chi}_{n},\vec{\chi}'_n}&=&\langle\vec{\chi}_n|\sum^N_{j=1}S^z_jS^z_{j+1}|\vec{\chi}'_{n}\rangle.
\end{eqnarray}
For the homogeneous XX ring described by $H_{\mathrm{XX}}$, it is shown in Ref.~\cite{SOME2018} that $F_{j;\vec{\eta}_{n+1},\vec{\chi}_n}\equiv\langle\vec{\chi}_n| S^-_j|\vec{\eta}_{n+1}\rangle$ admits a simple factorized form,
\begin{eqnarray}\label{Fjetachi}
F_{j;\vec{\eta}_{n+1},\vec{\chi}_n}=\frac{1}{\sqrt{N}}\left(\frac{2}{N}\right)^n e^{i(j-n)\Delta_{\vec{\eta}_{n+1},\vec{\chi}_n}}h_{\vec{\eta}_{n+1},\vec{\chi}_n},
\end{eqnarray}
where $\Delta_{\vec{\eta}_{n+1},\vec{\chi}_n}=\sum^{n+1}_{j=1}K^{(\sigma_{n+1})}_{\eta_j}-\sum^n_{i=1}K^{(\sigma_n)}_{\chi_i}$ is the momentum transfer between $|\vec{\eta}_{n+1}\rangle$ and $|\vec{\chi}_n\rangle$ and
\begin{eqnarray}\label{hfunction}
&& h_{\vec{\eta}_{n+1},\vec{\chi}_n}=\nonumber\\
&&\frac{\prod_{i>i'}(e^{-iK^{(\sigma_n)}_{\chi_i}}-e^{-iK^{(\sigma_n)}_{\chi_{i'}}})\prod_{j>j'}(e^{iK^{(\sigma_{n+1})}_{\eta_j}}-e^{iK^{(\sigma_{n+1})}_{\eta_{j'}}})}{\prod^n_{i=1}\prod^{n+1}_{j=1}(1-e^{-i(K^{(\sigma_{n+1})}_{\eta_j}-K^{(\sigma_n)}_{\chi_i})})}\nonumber\\
\end{eqnarray}
is a function of the momenta~\cite{Hanamura}. From Eq.~(\ref{Fjetachi}), we immediately get
\begin{eqnarray}
&&F_{\vec{\eta}_{n+1},\vec{\chi}_n}(\{g_j\})\nonumber\\
&=&\left(\frac{2}{N}\right)^n\frac{\tilde{g}_{\Delta_{\vec{\eta}_{n+1},\vec{\chi}_n}}e^{-in\Delta_{\vec{\eta}_{n+1},\vec{\chi}_n}}}{\sqrt{N}}h_{\vec{\eta}_{n+1},\vec{\chi}_n},
\end{eqnarray}
where $\tilde{g}_q=\sum^N_{j=1}e^{iqj}g_j$ is the Fourier transform of $\{g_j\}$. Using $S^z_j=\frac{1}{2}-S^-_jS^+_j$, we similarly obtain
\begin{eqnarray}\label{Gfun}
&& G_{\vec{\chi}_n,\vec{\chi}'_n}(\{g'_j\})= \frac{1}{2}\delta_{\vec{\chi}_n,\vec{\chi}'_n}\sum_jg'_j\nonumber\\
&&~~~-\left(\frac{2}{N}\right)^{2n}\frac{\tilde{g}'^*_{\Delta_{\vec{\chi}_n,\vec{\chi}'_n}}e^{in\Delta_{\vec{\chi}_n,\vec{\chi}'_n}}}{N}\bar{h}_{\vec{\chi}_n,\vec{\chi}'_n},
\end{eqnarray}
where
\begin{eqnarray}
\bar{h}_{\vec{\chi}_n,\vec{\chi}'_n}&=&\sum_{\vec{\eta}_{n+1}}h_{\vec{\eta}_{n+1},\vec{\chi}_n}h^*_{\vec{\eta}_{n+1},\vec{\chi}'_n}.
\end{eqnarray}
As a byproduct, the matrix elements of the staggered magnetization,
\begin{eqnarray}
m_s\equiv\frac{1}{N}\sum^N_{j=1}(-1)^jS^z_j,
\end{eqnarray}
which measures the antiferromagnetic order in the XXZ chain with $J'/J>0$, can be obtained by setting $g'_j=\frac{1}{N}e^{i\pi j}$ in Eq.~(\ref{Gfun}):
\begin{eqnarray}
&& m_{s;\vec{\chi}_n,\vec{\chi}'_n}=(-1)^{n-1}\left(\frac{2}{N}\right)^{2n}\frac{\delta(\Delta_{\vec{\chi}_n,\vec{\chi}'_n},\pi)}{N}\bar{h}_{\vec{\chi}_n,\vec{\chi}'_n},\nonumber\\
\end{eqnarray}
where
\begin{eqnarray}
\delta(x,y)=\begin{cases}
    1,  & \text{$x-y=2\pi m,~m\in Z$},\\
    0, & \mathrm{otherwise}.
  \end{cases}
\end{eqnarray}
Finally, the matrix elements $\bar{G}_{\vec{\chi}_{n},\vec{\chi}'_n}$ can also be calculated from Eq.~(\ref{Fjetachi}) and has the form
\begin{eqnarray}\label{GB}
&&\bar{G}_{\vec{\chi}_{n},\vec{\chi}'_n}=\left(n -\frac{3 N}{4}\right)\delta_{\vec{\chi}_n,\vec{\chi}'_n}\nonumber\\
&+&\left(\frac{2}{N}\right)^{4n}\frac{\delta(\Delta_{\vec{\chi}_n,\vec{\chi}'_n},0)}{N}\sum_{\vec{\eta}_n}e^{ i\Delta_{\vec{\chi}_n,\vec{\eta}_n}} \bar{h}_{\vec{\chi}_n,\vec{\eta}_n}\bar{h}_{\vec{\eta}_n,\vec{\chi}'_n}.\nonumber\\
\end{eqnarray}
\par The advantage of using the eigenbasis of the XX chain now becomes clear: the system-bath coupling constants simply enter the matrix elements $F_{\vec{\eta}_{n+1},\vec{\chi}_n}(\{g_j\})$ and $G_{\vec{\chi}_{n},\vec{\chi}'_n}(\{g'_j\})$ through the Fourier transforms $\tilde{g}_{\Delta_{\vec{\eta}_{n+1},\vec{\chi}_n}}$ and $\tilde{g}'^*_{\Delta_{\vec{\chi}_n,\vec{\chi}'_n}}$. The main task is to calculate the function $h_{\vec{\eta}_{n+1},\vec{\chi}_n}$ given by Eq.~(\ref{hfunction}). Moreover, the matrix elements $\bar{G}_{\vec{\chi}_{n},\vec{\chi}'_n} $given by Eq.~(\ref{GB}) also provide an alternative way to diagonalize the XXZ chain in a basis where $H_{\mathrm{XX}}$ is diagonal (in contrast, the Ising term $H_{\mathrm{Z}}$ is diagonal in the real basis formed by the Ising configurations). In passing we mention that, in principle, the dynamics of the generalized Heisenberg star can also be accurately simulated by using the Chebyshev expansion technique~\cite{PRB2016,CETPRE,Anders}.
\subsection{Initial states, time-evolved states, and equations of motion}\label{EOMapp}
\par We assume a separable initial state for the star,
\begin{eqnarray}
|\psi(0)\rangle&=&|\phi^{(\mathrm{S})}\rangle\otimes|\phi^{(\mathrm{B})}\rangle,
\end{eqnarray}
where $|\phi^{(\mathrm{S})}\rangle$ is a general pure state of the central spin,
\begin{eqnarray}
|\phi^{(\mathrm{S})}\rangle=\sum^{S}_{s_z=-S}a_{s_z}|s_z\rangle,
\end{eqnarray}
with $\sum^{S}_{s_z=-S}|a_{s_z}|^2=1$. Similarly, $|\phi^{(\mathrm{B})}\rangle$ is a pure state of the XXZ bath and can generally be written as a linear combination of the component states having fixed number of fermionic excitations:
\begin{eqnarray}\label{phiB}
|\phi^{(\mathrm{B})}\rangle=\sum^N_{n=0}\sum_{\vec{\eta}_n}b_{\vec{\eta}_n}|\vec{\eta}_n\rangle,
\end{eqnarray}
where $\sum^N_{n=0}\sum_{\vec{\eta}_n}|b_{\vec{\eta}_n}|^2=1$. Since the time evolution occurs in each sector with fixed $\alpha$, the most general form of the time-evolved state is
\begin{eqnarray}\label{psit}
|\psi(t)\rangle&=&|\psi^{\mathrm{(I)}}(t)\rangle+|\psi^{\mathrm{(II)}}(t)\rangle+|\psi^{\mathrm{(III)}}(t)\rangle.
\end{eqnarray}
According to the classification of different structures of the magnetization sectors listed in Appendix~\ref{AppA}, the three parts of the time-evolved state read
\begin{eqnarray}\label{psitI}
|\psi^{\mathrm{(I)}}(t)\rangle&=& \sum^{S}_{\alpha=-S}\sum^{\alpha+S}_{n=0}\sum_{\vec{\eta}_n}A^{\mathrm{I},\alpha}_{\alpha-n,\vec{\eta}_n} |\alpha-n\rangle|\vec{\eta}_n\rangle,\nonumber\\
|\psi^{\mathrm{(II)}}(t)\rangle&=&\sum^{N-S-1}_{\alpha=S+1}\sum^{\alpha+S}_{n=\alpha-S}\sum_{\vec{\eta}_n}A^{\mathrm{II},\alpha}_{\alpha-n,\vec{\eta}_n}|\alpha-n\rangle|\vec{\eta}_n\rangle\nonumber\\
|\psi^{\mathrm{(III)}}(t)\rangle&=&\sum^{N+S}_{\alpha=N-S}\sum^{N}_{n=\alpha-S}\sum_{\vec{\eta}_n}A^{\mathrm{III},\alpha}_{\alpha-n,\vec{\eta}_n}|\alpha-n\rangle|\vec{\eta}_n\rangle,\nonumber\\
\end{eqnarray}
with initial conditions
\begin{eqnarray}
A^{\mathrm{i},\alpha}_{\alpha-n,\vec{\eta}_n}=a_{\alpha-n}b_{\vec{\eta}_n},~\mathrm{i=I,II,III}.
\end{eqnarray}
\begin{figure}
\includegraphics[width=.76\textwidth]{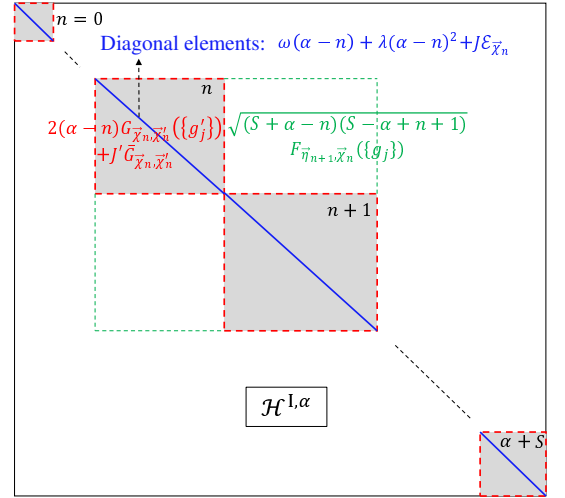}
\caption{Structure of the block Hamiltonian $\mathcal{H}^{\mathrm{I},\alpha}$ for a fixed $\alpha$ with $-S\leq \alpha\leq S$. The main diagonal blocks are associated with the spin-operator matrix elements $G_{\vec{\chi}_{n},\vec{\chi}'_n}(\{g'_j\})$ and $\bar{G}_{\vec{\chi}_{n},\vec{\chi}'_n}$ (with red sides), while the off-diagonal blocks are associated with the matrix elements $F_{\vec{\eta}_{n+1},\vec{\chi}_n}(\{g_j\})$ (with dotted green sides). The blue line indicates the diagonal terms.}
\label{Fig2}
\end{figure}
\par For a fixed $\alpha$ satisfying $-S\leq \alpha\leq S$, let
\begin{eqnarray}
\vec{A}^{\mathrm{I},\alpha}=(A^{\mathrm{I},\alpha}_{\alpha,\vec{\eta}_0},\{A^{\mathrm{I},\alpha}_{\alpha-1,\vec{\eta}_1}\},\cdots,\{A^{\mathrm{I},\alpha}_{-S,\vec{\eta}_{\alpha+S}}\})^{\mathrm{T}}\nonumber
\end{eqnarray}
be the amplitude vector in the ordered basis
\begin{eqnarray}
|\alpha\rangle|0\rangle,|\alpha-1\rangle\{|\vec{\eta}_1\rangle\},\cdots,|-S\rangle\{|\vec{\eta}_{\alpha+S}\rangle\},\nonumber
\end{eqnarray}
the equations of motion of $\vec{A}^{\mathrm{I},\alpha}$ then read
\begin{eqnarray}
i\frac{d}{dt}\vec{A}^{\mathrm{I},\alpha}=\mathcal{H}^{\mathrm{I},\alpha}\vec{A}^{\mathrm{I},\alpha},
\end{eqnarray}
where $\mathcal{H}^{\mathrm{I},\alpha}$ is a $D^{\mathrm{I},\alpha}\times D^{\mathrm{I},\alpha}$ matrix with $D^{\mathrm{I},\alpha}=\sum^{\alpha+S}_{n=0}\binom{N}{n}$. The structure of $\mathcal{H}^{\mathrm{I},\alpha}$ is shown in Fig.~\ref{Fig2}. Similar analysis can be made for categories $\mathrm{II}$ and $\mathrm{III}$. To obtain the time-evolved state $|\psi(t)\rangle$, we need only to simulate the time evolution of each amplitude vector $\vec{A}^{\mathrm{i},\alpha}$ governed by $\mathcal{H}^{\mathrm{i},\alpha}$ in each subspace with fixed $\alpha$. In our numerical simulations this is achieved through an exact diagonalization of the matrix $\mathcal{H}^{\mathrm{i},\alpha}$.
\par In the following, we will apply our method to study the dynamics of the system starting with two different bath initial states, namely the N\'eel state $|\mathrm{AF}\rangle=|\downarrow\uparrow\cdots\downarrow\uparrow\rangle$ and the ground state $|G_{\mathrm{XXZ}}\rangle$ of the XXZ bath. In the former case, we mainly focus on the influence of system-bath coupling on the relaxation of antiferromagnetic order within the XXZ bath; while in the latter case we are interested in the effects of internal phases of the bath on the reduced dynamics of the central spin.
\section{The N\'eel state $|\mathrm{AF}\rangle$: relaxation of antiferromagnetic order within the XXZ bath}\label{SecIII}
\par The N\'eel state $|\mathrm{AF}\rangle=|\downarrow\uparrow\cdots\downarrow\uparrow\rangle$ is one of the two degenerate ground states of the XXZ chain in the Ising limit $J'/J\to\infty$. It has been employed to detect the relaxation of antiferromagnetic order in the XXZ chain after a quantum quench~\cite{PRL2009,NJP2010}, to study the decoherence dynamics of a qubit coupled to both noninteracting~\cite{PRL2013} and interacting~\cite{PRB2016} spin baths, to make the connection between dynamical quantum phase transitions and order parameter dynamics~\cite{Heyl2014}, and more recently, to probe information scrambling in integrable and nonintegrable spin chain models~\cite{WVLiu2020,Cala2020}. Moreover, the N\'eel state lives in the largest magnetization sector (of dimension $\binom{N}{N/2}$) of the pure XXZ chain and could lead to nontrivial real-time dynamics.
\subsection{Without system-bath coupling: $g_j=g'_j=0$}
\par
\begin{figure}
\includegraphics[width=.60\textwidth]{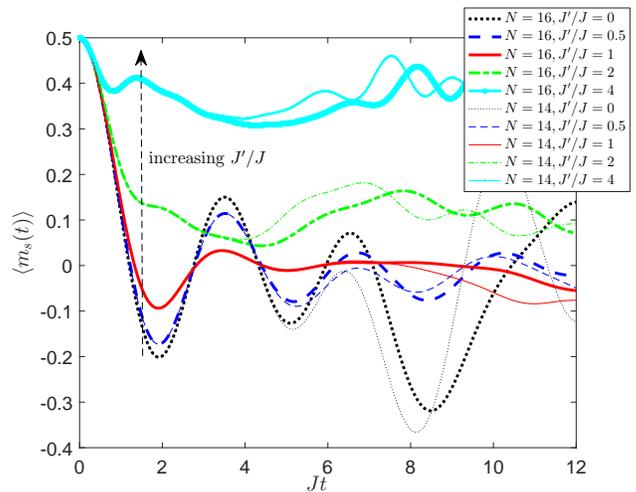}
\caption{Dynamics of the staggered magnetization $\langle m_s(t)\rangle$ in the pure XXZ chain ($g_j=g'_j=0,~\forall j$) with $N=14$ (thin curves) and $16$ (thick curves) sites. The initial state is chosen as the N\'eel state $|\mathrm{AF}\rangle=|\downarrow\uparrow\cdots\downarrow\uparrow\rangle$. The time evolution up to $Jt=4$ is independent of $N$ and expected to match the thermodynamic limit result. }
\label{FigpureXXZ}
\end{figure}

 To show the validity of our method, we first calculate the dynamics of the staggered magnetization $\langle m_s(t)\rangle$ in a pure antiferromagnetic XXZ chain (without system-bath coupling), which has been thoroughly studied in Refs.~\cite{PRL2009,NJP2010} by using the infinite-size time-evolving block decimation algorithm. In Fig.~\ref{FigpureXXZ} we plot $\langle m_s(t)\rangle$ for both $N=14$ (thin lines) and $N=16$ (thick lines). It can be seen that the short- to intermediate-time dynamics of $\langle m_s(t)\rangle$ ($Jt<4$) is insensitive to the variation of $N$ and is expected to faithfully capture the result in the thermodynamic limit $N\to\infty$. Actually, these features were numerically confirmed in larger chains up to $N=24$ by using exact diagonalization based on a Lanczos algorithm~\cite{Heyl2014}. Similar size-independent short-time dynamical behaviors of other order parameters in other spin models, e.g., the transverse/longitudinal magnetization dynamics in the quantum Ising chain~\cite{PRE2020,Rossini2020,PRB2021}, were also observed.
\par For large enough chains and at long times, it is found in Refs.~\cite{PRL2009,NJP2010} that $\langle m_s(t)\rangle$ exhibits an oscillatory (a nonoscillatory) decay for $0<J'/J<1$ ($J'/J>1$), while the fastest relaxation occurs close to the critical point $J'/J=1$. For $N=16$, although we observe a rough relaxation of $\langle m_s(t)\rangle$ (red solid curve, $J'/J=1$) around $Jt=7.5$, a deviation from $\langle m_s(t)\rangle\approx 0$ occurs at later times, and $\langle m_s(t)\rangle$ starts to oscillate at long times [see Fig.~\ref{Figmst}(a) below]. This is mainly due to the finiteness of the relevant Hilbert space (of dimension $\binom{16}{8}=12870$ for $N=16$). We now ask the question: How does the interaction between the central spin and the XXZ bath alter the dynamical behaviors of the antiferromagnetic order within the chain? As we will see, both the system-bath coupling strength and the size of the central spin have significant effects on the initial decay and long-time dynamics of $\langle m_s(t)\rangle$.
\subsection{Including the system-bath coupling}
\par  We use the following inhomogeneous system-bath coupling~\cite{PRL2013}
\begin{eqnarray}\label{gj}
g_j=g'_j/\Lambda=\frac{g}{N}e^{-\frac{j-1}{N}},
\end{eqnarray}
which corresponds to a Gaussian wave function in a two-dimensional quantum dot~\cite{PRB2004}. The initial state of the central spin is chosen as an equally weighted state
\begin{eqnarray}\label{PsiS}
|\phi^{(\mathrm{S})}\rangle=\frac{1}{\sqrt{2S+1}}(|S\rangle+|S-1\rangle+\cdots+|-S\rangle),
\end{eqnarray}
and the initial state of the XXZ bath is assumed to be the N\'eel state $|\phi^{(\mathrm{B})}\rangle=|\mathrm{AF}\rangle$. The parameter $g$ appearing in Eq.~(\ref{gj}) defines the overall energy scale through the relation~\cite{PRB2016}
\begin{eqnarray}
\omega_{\mathrm{fluc}}=\frac{2g}{N}e^{\frac{1}{N}-1}\sqrt{\frac{e^2-1}{e^{\frac{2}{N}}-1}}.
\end{eqnarray}
\begin{figure*}
\includegraphics[width=0.98\textwidth]{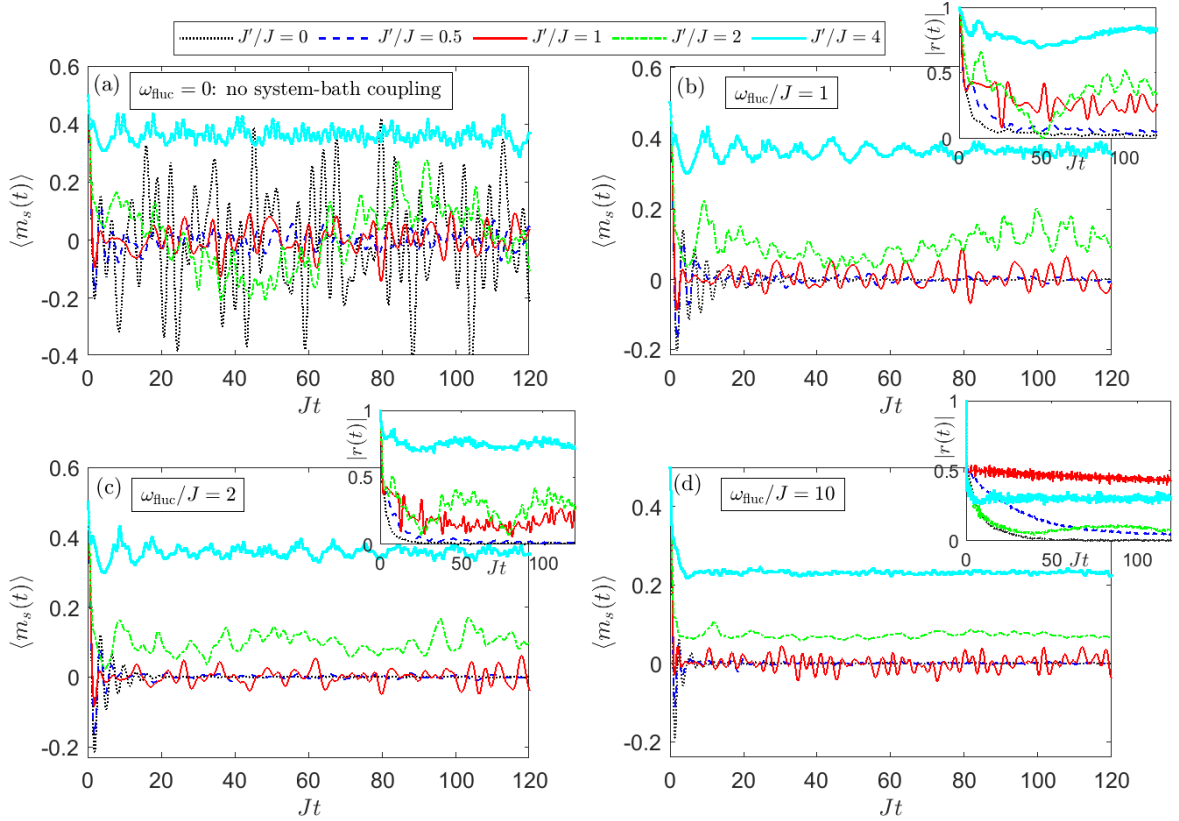}
\caption{Time evolution of the staggered magnetization $\langle m_s(t)\rangle$ in a generalized Heisenberg star composed of a $S=1/2$ central spin and an XXZ chain with $N=16$ sites. The XXZ bath is prepared in the N\'eel state $|\mathrm{AF}\rangle=|\downarrow\uparrow\cdots\downarrow\uparrow\rangle$ and the initial state of the central spin is $|\phi^{(\mathrm{S})}\rangle=\frac{1}{2}(|\uparrow\rangle+|\downarrow\rangle)$. The corresponding dynamics of the central spin decoherence factor $|r(t)|=|\langle S_+(t)/\langle S_+(0)\rangle|$ is shown in the right upper corners of panels (b), (c) and (d). Other parameters: $\Lambda=1$ and $\omega=\lambda=0$.}
\label{Figmst}
\end{figure*}
Note that the Fourier transform of $g_j$ has a simple form,
\begin{eqnarray}
\tilde{g}_q&=&\frac{g}{N}  \frac{1-e^{(i qN-1)}}{e^{-iq}-e^{ -\frac{1}{N}}}.
\end{eqnarray}
\par In this section, we focus on the case of $S=1/2$. Since $|\mathrm{AF}\rangle$ lives in the manifold with excitation number $n=N/2$, the index $\alpha$ takes two possible values, $\alpha=(N\pm1)/2$, and the time-evolved state is of type II for $N>2$, as can be seen from Eq.~(\ref{psitI}). The dimension of the relevant Hilbert space is $2 (\binom{16}{8}+\binom{16}{7}) =48620$ for $N=16$, which is large enough to observe nontrivial dynamics and the simulations can be performed on a personal workstation.
\par To see the effects of the system-bath coupling on the internal dynamics of the bath, we first use the intrabath coupling $J$ rather than $\omega_{\mathrm{fluc}}$ as an overall energy scale. For comparison, in Fig.~\ref{Figmst}(a) we plot the time evolution of $\langle m_s(t)\rangle$ on a longer time scale up to $Jt=120$ in the pure XXZ chain with $N=16$ sites. As expected, for all values of $J'/J$ considered, $\langle m_s(t)\rangle$ does not relax in the long time limit but exhibit irregular oscillations due to the finite size effect. Since there is no direct interaction between the central spin and the XXZ bath, the former does not show any decoherence. Once the system-bath coupling is introduced, the uncorrelated initial state will become entangled and the decoherence of the central spin and the related relaxation of the aniferromagnetic order in the chain occur simultaneously.
\begin{figure}
\includegraphics[width=0.51\textwidth]{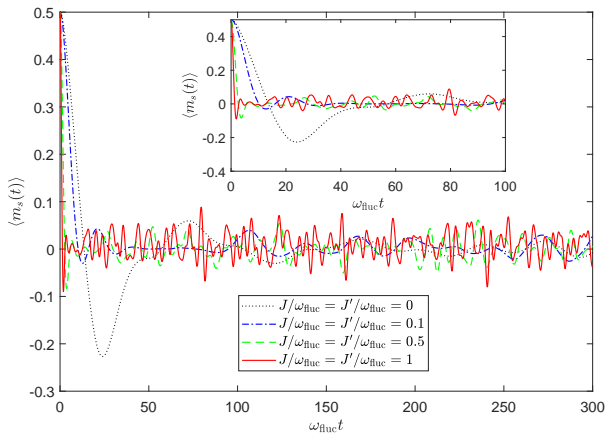}
\caption{Time evolution of the staggered magnetization $\langle m_s(t)\rangle$ in a Heisenberg star composed of a single $S=1/2$ central spin and an isotropic XXZ chain with $N=16$ sites and $J'/J=1$. The inset shows the dynamics up to an intermediate time $\omega_{\mathrm{fluc}}t=100$. Note that we use $\omega_{\mathrm{fluc}}$ as an overall energy scale so that the initial decay of $\langle m_s(t)\rangle$ is accelerated by increasing the intrabath coupling $J/\omega_{\mathrm{fluc}}$. Other parameters: $\Lambda=1$ and $\omega=\lambda=0$.}
\label{varyingJ}
\end{figure}
\begin{figure}
\includegraphics[width=0.52\textwidth]{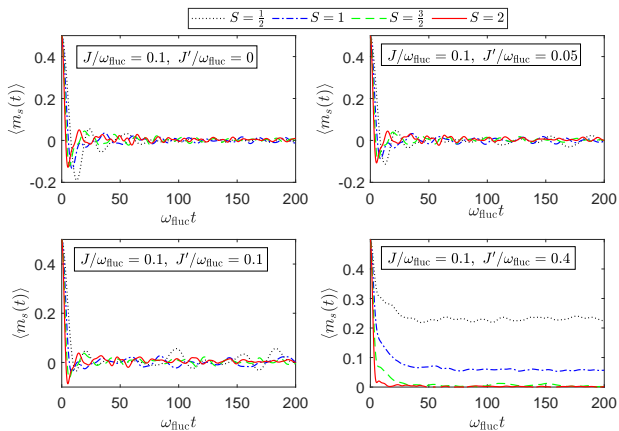}
\caption{Time evolution of the staggered magnetization $\langle m_s(t)\rangle$ in a Heisenberg star composed of an isotropic XXZ chain with $N=14$ and a central spin with $S=1/2$, $1$, $3/2$ and $2$. The intrabath coupling is chosen to be $J/\omega_{\mathrm{fluc}}=0.1$ (strong system-bath coupling) and the results for several values of the anisotropy parameter $J'/J$ are presented. Other parameters: $\Lambda=1$ and $\omega=\lambda=0$.}
\label{varyingS}
\end{figure}
\par Figure~\ref{Figmst}(b) shows $\langle m_s(t)\rangle$ in the weak system-bath coupling limit with $\omega_{\mathrm{fluc}}/J=1$, for which the largest hyperfine interaction is $2g/N\approx 0.37\omega_{\mathrm{fluc}}<J$ for $N=16$. At short times $\langle m_s(t)\rangle$ behaves similarly to the result without system-bath coupling [comparing to Fig.~\ref{Figmst}(a)]. In the long time limit, $\langle m_s(t)\rangle$ still exhibits oscillatory behaviors for $J'/J> 1$, but with positive amplitudes. At the critical point $J'/J=1$, $\langle m_s(t)\rangle$ oscillates around the zero value. Interestingly, for $J'/J<1$ we observe that $\langle m_s(t)\rangle$ relaxes to nearly zero after an initial oscillatory decay (black dotted and blue dashes curves for $J'/J=0$ and $0.5$, respectively). This is in sharp contrast to the case without system-bath coupling where $\langle m_s(t)\rangle$ oscillates intensively for $J'/J=0$ [Fig.~\ref{Figmst}(a)]. Therefore, even weak system-bath coupling can assist the long-time relaxation of the antiferromagnetic order for an intermediate-size (17 spins in total) generalized Heisenberg star with easy-plane anisotropy $0\leq J'/J<1$.
\par These behaviors persist for a stronger system-bath coupling with $\omega_{\mathrm{fluc}}/J=2$ [Fig.~\ref{Figmst}(c)], where we further observe that for $J'/J<1$ the initial oscillatory stage before the relaxation taking place becomes shorter. The situation is more interesting when we enter the strong system-bath coupling regime [Fig.~\ref{Figmst}(d)]. Except for the critical point $J'/J=1$ for which $\langle m_s(t)\rangle$ still experiences oscillations around its zero mean value, in all the other cases $\langle m_s(t)\rangle$ quickly approaches an almost steady value. The steady value of $\langle m_s(t)\rangle$ is nearly zero (large than zero) for $J'/J<1$ ($J'/J>1$).
\par Overall, the system-bath coupling has significant influence on the long-time dynamics of $\langle m_s(t)\rangle$, although it seems that the initial decay of $\langle m_s(t)\rangle$ is insensitive with respect to varying $\omega_{\mathrm{fluc}}/J$. Qualitatively, since the central spin and the XXZ bath are initially uncorrelated, at short times the magnetic order encoded in the antiferromagnetic bath state $|\mathrm{AF}\rangle$ only spreads within the chain through the nearest-neighbor intrabath coupling $J$. The system-bath coupling, which can be viewed as a kind of long-range interaction within the star, connects each bath spin with the common central spin and generates effective spin-spin couplings among the bath spins. It thus takes a longer period of time to establish the correlation between the two part and induce both the decoherence of the central spin and the relaxation of the antiferromagnetic order.
\par To further understand the above dynamical behaviors of $\langle m_s(t)\rangle$, we also plot in the right upper corners of Fig.~\ref{Figmst}(b)-(d) the corresponding decoherence factor $|r(t)|=|\langle S_+(t)/\langle S_+(0)\rangle|$~\cite{PRA2005} of the spin-1/2 central spin. We find that the relaxation of $\langle m_s(t)\rangle$ at long times for $J'/J<1$ is accompanied by the decay of $|r(t)|$. The sharp decay of $|r(t)|$ in the case of $J'/J=0$ has been demonstrated in Ref.~\cite{PRB2016} using the Chebyshev expansion technique in an $N=16$ XX chain. For $J'/J\geq 1$, $|r(t)|$ shows irregular oscillations around a finite value in a similar way as $\langle m_s(t)\rangle$. The central spin decoherence and the dynamics of the staggered magnetization within the bath are therefore in some sense correlated.
\par We can also look at the dynamics of the antiferromagnetic order from the opposite limit with vanishing intrabath coupling and fixed system-bath coupling. In this integrable limit, there is no direct interaction among the bath spins and the decay of $\langle m_s(t)\rangle$ is solely governed by the system-bath coupling. The central spin decoherence starting from $|\mathrm{AF}\rangle$ for such a noninteracting bath has been thoroughly studied in Ref.~\cite{PRL2013} for large baths using a combination of algebraic Bethe ansatz and Monte Carlo simulation. The black dotted curve in Fig.~\ref{varyingJ} shows the evolution of $\langle m_s(t)\rangle$ for $J=J'=0$, where $\langle m_s(t)\rangle$ experiences a slower initial decay (see also the inset) and oscillates smoothly around zero at long times. Increasing the intrabath coupling to $J/\omega_{\mathrm{fluc}}=J'/\omega_{\mathrm{fluc}}=0.1$ induces a faster initial decay and an increase in the long-time oscillation frequency. For all the intrabath couplings considered, the initial decay rate (the time at which the first minimum of $\langle m_s(t)\rangle$ reaches) increases (decreases) with increasing $J/\omega_{\mathrm{fluc}}$. These behaviors can also be qualitatively understood from the fact that the intrabath coupling generates spin flips between two nearest-neighbor bath spins which tend to destroy the initial antiferromagnetic order on a short time scale.
\par Before ending this section, we finally discuss the effect of the central spin size $S$ on the dynamics of the antiferromagnetic order. To this end, we focus on the strong system-bath limit with $J/\omega_{\mathrm{fluc}}=0.1$. Due to the limitation of the computation resources, we present in Fig.~\ref{varyingS} the time evolution of $\langle m_s(t)\rangle$ for $N=14$ and $S=1/2,~3/2,~1$, and $2$. For $J'/J<1$, increasing $S$ for fixed $J'/J$ generally accelerates the initial decay of $\langle m_s(t)\rangle$, rendering the first minimum of $\langle m_s(t)\rangle$ to be reached earlier. Moreover, the oscillation amplitude at long times is suppressed when $S$ increases. Interestingly, for $J'/J>1$  (right bottom panel) we find that a larger $S$ results in a lower steady value of $\langle m_s(t)\rangle$ and for $S=2$ the staggered magnetization nearly vanishes at long times. Actually, for fixed $S<2N$ the index $\alpha$ can take $2S+1$ possible values, i.e., $N/2-S,N/2-S+1,\cdots,N/2+S$ for the equally weighted state $|\phi^{(\mathrm{S})}\rangle$. As a result, there are effectively $2S+1$ channels for the XXZ bath to interact with the central spin, inducing a faster relaxation of the antiferromagnetic order.
\section{The ground state $|G_{\mathrm{XXZ}}\rangle$: Growth of the R\'enyi entanglement entropy of the central spin}\label{SecIV}
\par In this section, we study the dynamics of the central spin when the XXZ bath is prepared in its ground state $|\phi^{(\mathrm{B})}\rangle=|G_{\mathrm{XXZ}}\rangle$ for some fixed $J'/J$. The initial state of the central spin is still assumed to be an equally weighted superposition state $|\phi^{(\mathrm{S})}\rangle$ given by Eq.~(\ref{PsiS}). The dynamical protocol can be considered as a sudden quench in the hyperfine coupling strength: at $t=0^-$ the whole system lies in a separable state associated with $g_j=g'_j=0,~\forall j$, and then one suddenly turns on the system-bath coupling with strengths given by Eq.~(\ref{gj}).
\par It is known that for even $N$ and $J'/J>-1$ the ground state of $H_{\mathrm{B}}$ is nondegenerate and possesses magnetization $l_z=0$; while for $J'/J<-1$ the bath is ferromagnetic and has two degenerate fully polarized ground states $|\uparrow\cdots\uparrow\rangle$ and $|\downarrow\cdots\downarrow\rangle$~\cite{Affleck1986}. Below we focus on the case of  $J'/J>0$, so that the initial bath state $|\phi^{(\mathrm{B})}\rangle=\sum_{\vec{\eta}_{\frac{N}{2}}}b_{\vec{\eta}_{\frac{N}{2}}}|\vec{\eta}_{\frac{N}{2}}\rangle$ lives in the subspace with $n=N/2$. The coefficients $\{b_{\vec{\eta}_{\frac{N}{2}}}\}$ can be determined numerically by solving an
eigenvalue problem of $H_{\mathrm{B}}$ in the $l_z=0$ sector and finding the lowest energy state.
\par We first note that in the R-V model a spin-1/2 central spin does not experience any decoherence from the initial state $|\phi^{(\mathrm{S})}\rangle\otimes|\phi^{(\mathrm{B})}\rangle$. Actually, the absolute ground state of an XXX chain is a singlet state $|G_{\mathrm{XXX}}\rangle$ having angular momentum $l=0$~\cite{Lieb1961}, so that the total angular momentum of the system must be $\mathcal{J}=1/2$. It is easy to see from Eq.~(\ref{RV2}) that the initial state is an eigenstate of $H_{\mathrm{R-V}}$ with eigenenergy $E^{(\mathrm{g})}_{\mathrm{B}}$ and hence only acquires a phase factor during the time evolution, where $E^{(\mathrm{g})}_{\mathrm{B}}$ is the ground state energy of the XXX chain. However, the time evolution becomes nontrivial when one goes beyond the Richter-Voigt point.
\par In this subsection, we mainly focus on the case of $S=1$, for which the reduced density matrix of the central spin has a closed form~\cite{QMbook}
\begin{eqnarray}
\rho_{11}&=&1+\frac{1}{2}(a_z-q_{xx}-q_{yy}),\nonumber\\
\rho_{22}&=&-1+q_{xx}+q_{yy},\nonumber\\
\rho_{33}&=&1-\rho_{11}-\rho_{22},\nonumber\\
\rho_{12}&=& \rho^*_{21}=\frac{1}{2\sqrt{2}}[a_x+q_{zx}-i(a_y+q_{yz})],\nonumber\\
\rho_{13}&=& \rho^*_{31}=\frac{1}{2}(q_{xx}- q_{yy}-iq_{xy}),\nonumber\\
\rho_{23}&=&\rho^*_{32}=\frac{1}{2\sqrt{2}}[a_x-q_{zx}-i(a_y-q_{yz})],
\end{eqnarray}
where
\begin{eqnarray}
a_i&=&\langle S_i\rangle,~i=x,y,z,\nonumber\\
q_{ii}&=&\langle(S_i)^2\rangle,~i=x,y,\nonumber\\
q_{ij}&=&\langle S_i S_j +S_j S_i \rangle,~ij=xy,yz,zx.
\end{eqnarray}
\par We are interested in the time evolution of the R\'enyi entanglement entropy of the central spin ($\alpha\ge0$ and $\alpha\neq 1$)
\begin{eqnarray}
R_\alpha(\rho_{\mathrm{CS}})=\frac{1}{1-\alpha}\ln \mathrm{Tr}(\rho^\alpha_{\mathrm{CS}}).
\end{eqnarray}
The R\'enyi entanglement entropy naturally generalizes the von Neumann entropy and reduces to the latter when $\alpha\to 1$. The R\'enyi entanglement entropy for $\alpha=2$ has been measured in cold-atom experiments~\cite{RSexp}.
\begin{figure}
\includegraphics[width=.52\textwidth]{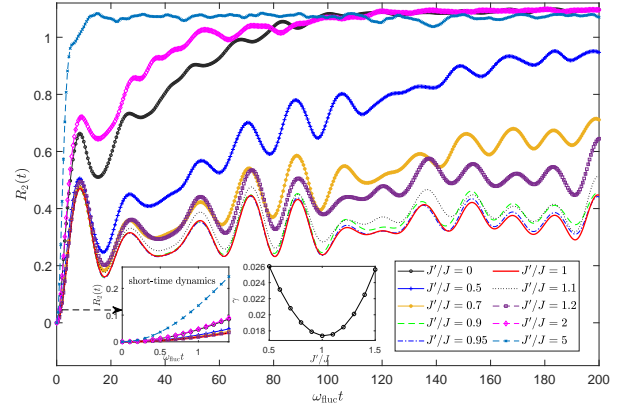}
\caption{Dynamics of the R\'enyi entanglement entropy $R_2(t)$ of a spin-$1$ central spin coupled to an XXZ chain with $N=14$ sites. The bath is prepared in the ground state $|\phi^{(\mathrm{B})}\rangle=|G_{\mathrm{XXZ}}\rangle$ for fixed $J'/J$ and the intrabath coupling is chosen as $J/\omega_{\mathrm{fluc}}=0.1$ (strong system-bath coupling).  The central spin is prepared in an equally weighted superposition state $|\phi^{(\mathrm{S})}\rangle=\frac{1}{\sqrt{3}}(|1\rangle+|0\rangle+|-1\rangle)$. The left inset shows the short-time dynamics of $R_2(t)$ up to $\omega_{\mathrm{fluc}}t=1.5$ and the right inset shows the growth rate $\gamma$ [see Eq.~(\ref{decay})] as a function of $J'/J$.  Other parameters: $\Lambda=1$ and $\omega=\lambda=0$.}
\label{Fig5}
\end{figure}
\par Figure~\ref{Fig5} shows $R_2(t)$ after a sudden quench to the strong system-bath coupling regime with $J/\omega_{\mathrm{fluc}}=0.1$ for an XXZ chain with $N=14$ sites. The results for various values of $J'/J$ are shown to see the influence of different quantum phases of the XXZ chain on the dynamics of the R\'enyi entanglement entropy. In the limit of $J'=0$, the XXZ chain is reduced to the XX chain whose ground state is a fermionic Fock state $|\vec{\eta}_7\rangle=|1,2,3,4,12,13,14\rangle$ for $J>0$. In this case, $R_2(t)$ increases rapidly at short times and gradually approaches its maximal value $\sim \ln 3$ at long times after experiencing oscillations in the intermediate-time regime. The overall profile of $R_2(t)$ is found to be lifted down as $J'/J$ increases from $0$ to $1$ within the gapless phase. Remarkably, we observe that $R_2(t)$ acquires the lowest values at the critical point $J'/J=1$ (thick red curve), beyond which its magnitude increases again as $J'/J$ increases further in the gapped phase. Specially, in the large $J'/J$ limit $R_2(t)$ increases more abruptly at the beginning and approaches a steady value close to the maximal value $\ln 3$. Such a fast growth of the R\'enyi entanglement entropy might be an indicator of fast information scrambling since the system-bath couplings are essentially long-range interactions~\cite{WVLiu2020}.
\par It is also interesting to analyze the short-time dynamics of $R_2(t)$. Previous studies revealed that the short-time evolution of typical observables of the central spin is often of a Gaussian form~\cite{PRA2007,PRA2014,PRB2016}. The left inset of Fig~\ref{Fig5} displays $R_2(t)$ up to $\omega_{\mathrm{fluc}}t=1.5$. It can be seen that $R_2(t)$ indeed increases as
\begin{eqnarray}\label{decay}
R_2(t)\sim 1-e^{-\gamma(\omega_{\mathrm{fluc}}t)^2},
\end{eqnarray}
where $\gamma$ is a growth rate depending on the value of $J'/J$. The short-time behavior of $R_2(t)$ faithfully reflects its overall profile over long times. In the right inset of Fig~\ref{Fig5} we plot $\gamma$ as a function of $J'/J$. We find that $\gamma$ depends nonmonotonically on $J'/J$ and reaches a minimum at the critical point $J'/J=1$. In principle, the short-time dynamics for $\omega_{\mathrm{fluc}}t\ll 1$ can be captured through time-dependent perturbation analysis~\cite{PRA2014}. The details of the overall dynamics, however, depends on the total Hamiltonian and might be qualitatively explained based on a spectral analysis~\cite{NJP2010} of the effective Hamiltonians $\mathcal{H}^{\mathrm{II},\alpha}$. Although it is not straightforward to perform these analysis due to the complexity of the full quantum dynamics, our numerical results do reveal that interesting dynamics of both the staggered magnetization and the central spin can take place close to the critical point $J'/J=1$.
\section{Conclusions and Discussions}\label{SecV}
\label{sec-final}
\par In this work, we obtain exact dynamics of a generalized Heisenberg star made up of a spin-$S$ central spin and an inhomogeneously coupled antiferromagnetic XXZ chain. Such an interacting central spin model can be viewed either as a generalization of the homogeneous Heisenberg star studied by Richter and Voigt~\cite{Heisenbergstar} to the case of inhomogeneous system-bath coupling, or as an extension of the XX star~\cite{PRB2016} by including the Ising part of the nearest-neighbor intrabath coupling. The generalized Heisenberg star may be simulated in cold-atom systems and are relevant to molecular aggregates located in a cavity.
\par In contrast to previous studies in which the reduced dynamics of the central spin is mainly concerned, we focus on the influence of the central spin on the internal dynamics of the many-body spin bath. Based on the conservation of the total magnetization and using analytical expressions of spin-operator matrix elements in the uniform XX chain~\cite{SOME2018}, we calculate the time evolution of the antiferromagnetic order in the XXZ bath that is initially prepared in a N\'eel state. We find that even weak system-bath coupling can lead to nearly perfect relaxation of the staggered magnetization in the gapless phase with $0\leq J'/J<1$. This is in contrast to the case of vanishing system-bath coupling~\cite{PRL2009} where intensive oscillations of the staggered magnetization are observed due to the finite-size effect. In the gapped phase of the XXZ bath with $J'/J>1$, we find that the staggered magnetization decays rapidly at short times and approaches a positive steady value which increases with increasing $J'/J$. However, at the critical point $J'/J=1$, the oscillatory behavior of the staggered magnetization remains even at long times and for strong system-bath couplings. We also investigate the effect of the size of the central spin $S$ on the antiferromagnetic order relaxation. It is found that increasing $S$ not only accelerates the initial decay but also suppresses the steady value of the staggered magnetization in the gapped phase. These observations may stimulate further studies of the influence of simple systems on the nonequilibrium dynamics of the coupled many-body system.
\par We then turn to study the reduced dynamics of the central spin in the usual way for an XXZ bath prepared in its ground state. We focus on the dynamics of the R\'enyi entanglement entropy of a higher central spin with $S=1$. We find that the second order R\'enyi entanglement entropy $R_2(t)$ of the central spin acquires its lowest value at the critical point $J'/J=1$. By analyzing the short-time behavior of $R_2(t)$, we find that the R\'enyi entanglement entropy grows according to a Gaussian form with a growth rate depending nonmonotonically on increasing $J'/J$. Remarkably, the growth rate is also found to reach a minimum at the critical point $J'/J=1$. These results point out a possibility of detecting the critical behavior of quantum critical spin baths by probing the entanglement dynamics of the coupled central spin.\\
\\
\noindent{\bf Acknowledgements:}
N.W. thanks X.-W. Guan, H. Katsura, S.-W. Li, and J. Zou for useful discussions. This work was supported by the National Key R\&D Program of China under Grant No. 2021YFA1400803 and by the Natural Science Foundation of China (NSFC) under Grant No. 11705007.

\appendix
\section{Structure of the $M$-sectors for $S<\frac{N}{2}$}\label{AppA}
\par Starting with the state with the lowest magnetization $M=-S-\frac{N}{2}$ and denote $(s_z,l_z)$ as the configuration with fixed magnetizations for the central spin and the spin bath, then each $M$ corresponds to the following configurations:
\par $M=-S-\frac{N}{2}$: $(-S,-\frac{N}{2})$;
\par $M=-S-\frac{N}{2}+1$: $(-S,1-\frac{N}{2}),(-S+1,-\frac{N}{2})$;
\par $\vdots$
\par  $M=S-\frac{N}{2}$: $(-S,2S-\frac{N}{2}),\cdots,(S,-\frac{N}{2})$;
\par --------------------
\par  $M=S-\frac{N}{2}+1$: $(-S,2S-\frac{N}{2}+1),\cdots,(S,-\frac{N}{2}+1)$;
\par  $\vdots$
\par $M=-S+\frac{N}{2}-1$: $(-S, \frac{N}{2}-1),\cdots,(S,\frac{N}{2}-2S-1)$;
\par --------------------
\par $M=-S+\frac{N}{2}$: $(-S, \frac{N}{2}),\cdots,(S,\frac{N}{2}-2S)$;
\par $\vdots$
\par $M=S+\frac{N}{2}$: $(S,\frac{N}{2})$.
\par In summary, the $M$-sectors can be classified into three categories:
\par I) For $-S-\frac{N}{2}\leq M\leq S-\frac{N}{2}$, there are $S+\frac{N}{2}+1+M$ configurations of $(s_z,l_z)$ in each $M$-sector, among which $s_z$ can take values from $ M+\frac{N}{2}$ to $-S$, with the corresponding $l_z$ running from $-\frac{N}{2}$ to $M+S$. The dimension of this $M$-sector is $d_M=\sum^{M+S+\frac{N}{2}}_{j=0}\binom{N}{j}$.
\par II) For $S-\frac{N}{2}+1\leq M\leq-S+\frac{N}{2}-1$,  there are $2S+1$ configurations of $(s_z,l_z)$ in each $M$-sector, among which $s_z$ can take all the values from $S$ to $-S$, with the corresponding $l_z$ running from $M-S$ to $M+S$. The dimension of this $M$-sector is $d_M=\sum^{M+S+\frac{N}{2}}_{j=M-S+\frac{N}{2}}\binom{N}{j}$.
\par III) For $-S+\frac{N}{2}\leq M\leq S+\frac{N}{2}$, there are $1+S+\frac{N}{2}-M$ configurations of $(s_z,l_z)$ in each $M$-sector, among which $s_z$ can take values from $S$ to $M-\frac{N}{2}$, with the corresponding $l_z$ running from $ M-S+\frac{N}{2}$ to $N$. The dimension of this $M$-sector is $d_M=\sum^{N}_{j=M-S+\frac{N}{2}}\binom{N}{j}=\sum^{\frac{N}{2}-M+S}_{j= 0}\binom{N}{j}$.
\end{document}